# Origins of size effects in initially dislocation-free single-crystal metallic micro- and nanocubes

*Claire Griesbach[1], Seog-Jin Jeon[2], David Funes Rojas[3], Mauricio Ponga[3], Sadegh Yazdi[4], Siddhartha Pathak[5,6], Nathan Mara[5,7], Edwin L. Thomas[8,9], Ramathasan Thevamaran[1,\*]*

[1]*Department of Engineering Physics, University of Wisconsin, Madison, WI 53706.*

[2]*Department of Polymer Science and Engineering, Kumoh National Institute of Technology, Gumi, Gyoeongbuk 39177, South Korea.*

[3]*Department of Mechanical Engineering, University of British Columbia, Vancouver, BC, Canada.*

[4]*Renewable and Sustainable Energy Institute, University of Colorado-Boulder, Boulder, CO 80309, USA.*

[5]*Center for Integrated Nanotechnologies, Los Alamos and Sandia National Laboratories, Los Alamos, NM 87545.*

[6]*Department of Materials Science and Engineering, Iowa State University, Ames, IA 50011.*

[7]*Department of Chemical Engineering and Materials Science, University of Minnesota, Minneapolis, MN 55455-0132.*

[8]*Department of Materials Science and NanoEngineering, Rice University, Houston, TX 77005.*

[9]*Department of Materials Science and Engineering, Texas A and M University, College Station, TX 77843.*

[\*]Correspondence: thevamaran@wisc.edu (RT)

**Abstract**

We report phenomenal yield strengths—up to one-fourth of the theoretical strength of silver—recorded in microcompression testing of initially dislocation-free silver micro and nanocubes synthesized from a multistep seed-growth process. These high strengths and the massive strain bursts that occur upon yield are the results of initially dislocation-free single-crystal structure of the pristine samples that yield through spontaneous nucleation of dislocations. When the pristine samples are exposed to a focused ion-beam to fabricate pillars and then compressed, the dramatic strain burst does not occur, and they yield at a quarter of the strength of their pristine counterparts. Regardless of the defect-state of the samples prior to testing, a size effect is apparent—where the yield strength increases as the sample size decreases. Since dislocation starvation and single-arm-source mechanisms cannot explain a size effect on yield strength in dislocation-free samples, we investigate the dislocation nucleation mechanisms controlling the size effect through careful experimental observations and molecular statics simulations. We find that intrinsic or extrinsic symmetry breakers such as surface defects, edge roundness, external sample shape, or a high vacancy concentration can influence dislocation nucleation, and thus contribute to the size effect on yield strength in initially dislocation-free samples.

**1. Introduction**

Tailoring a material's structural features across length-scales enables broad control of the material's mechanical properties. For example, decreasing the grain size in polycrystalline metals can improve the strength of the material significantly—called Hall-Petch strengthening [1,2]. The effects of extrinsic sample size on the strength—when the sample size is reduced to the micrometer regime and below—have been studied more recently with the advent of nanomechanical testing devices. Performing *in situ* SEM microcompression tests on single-crystal micro- and nano-samples allows for investigating the extrinsic sample size effect without the influence of intrinsic grain size effect. Commonly, micro- and nano-pillars are fabricated from a bulk material using a focused ion beam (FIB) and then compressed using an *in situ* SEM nanoindenter [3]. Several studies using this method with various materials have shown an increase in strength with decreasing sample size [4–7]. Dou and Derby collected data from multiple tests done on FIB-fabricated face-centered-cubic (fcc) micropillars and found a power law which describes the size effect on strength: $\tau_y/G = 0.71(d/b)^{-0.66}$ [8], where $\tau_y$ is the shear yield strength, $G$ is the shear modulus, $d$ is the effective diameter of the sample, and $b$ is the Burger's vector.

The effects of FIB-induced artifacts, such as $Ga^+$ implantation and surface amorphization, on the nanomechanical response and size effects of single-crystals have been debated [9–11]. It has been shown that even small doses of $Ga^+$ ions (30keV, 50pA) used for imaging can create ~230-400 vacancies/$Ga^+$, while larger doses used for milling can even create dislocations that extend into the material [12]. However, several studies which used alternative methods to create the micropillars show a similar size effect as the FIB-fabricated pillars [13–15], suggesting that FIB-induced defects do not control the size effect in single-crystal samples with an initial dislocation concentration.

Two main mechanisms were proposed to explain the size effect on strength: dislocation starvation and the single-arm-source (SAS) theory [6]. The SAS theory suggests that the size effect originates from a decrease in size of single-arm-sources in smaller samples, which require higher levels of strength to activate. This mechanism has been shown to be more apparent in micron-sized samples [6,7,16–18]. Dislocation starvation [5] has been seen experimentally in nanometer sized samples [16,19,20]. In this mechanism, strengthening occurs because the pre-existing dislocations readily annihilate at the free surface and upon further compression, the dislocation-starved crystal needs to nucleate new dislocations at a higher activation energy [6]. Image forces drive the dislocations to the free surface and these are stronger if the dislocation is closer to the surface as is the case in smaller samples [21].

These mechanisms depend on pre-existing mobile dislocations; thus, they cannot adequately explain a size effect on yield strength in dislocation-free samples. Only a few studies have examined the size effect in dislocation-free single-crystal samples experimentally [20,22–24] because of the difficulty of producing dislocation-free samples from the nanometer to micrometer regime with usual microfabrication approaches. Large strain bursts are found to occur at high yield strengths and size effects are apparent [22–24]. Through *in situ* TEM compression testing, dislocations are seen nucleating from edges and surfaces [20]. Molecular dynamics (MD) simulations reveal that dislocations nucleate at areas of high stress concentration such as sharp edges or vertices [25,26]. Feruz and Mordehai suggest that the size effect is due to the higher nucleation threshold in smaller samples because of a mild and gradual stress gradient [25]. Experiments backed by MD simulations reveal that rounded edges and corners eliminate sites of stress concentration for heterogeneous nucleation, making homogenous nucleation more likely which requires higher levels of stress [23,24,27].

While most experiments of dislocation-free single-crystal samples reveal that there is a size effect on the yield strength, it is still not well-understood what causes this size effect. Several studies suggest that a truly defect-free single-crystal sample should yield at theoretical strength [23,28,29]. Here, we provide insights to this by investigating the deformation responses of initially dislocation-free single-crystal silver (Ag) nano- and microcubes ranging from 100-2000 nm. We also compare the deformation responses of FIB-milled pillars to the as-synthesized cubes. Both square and circular cross-sectional pillars are tested to investigate the influence of geometry on the size effect. Furthermore, we use molecular statics (MS) simulations to investigate the effects of cube edge and corner roundness and vacancy concentration on the size effect.

## 2. Material and Methods

*2.1 Silver as the model material*

We use Ag nano- and micro-cubes as the model material to investigate the size effect. Ag is a face-centered-cubic metal (space group: Fm$\bar{3}$m, No. 225) with high thermal and electrical conductivities, diamagnetism, high reflectance, and low emittance. It is often used in superconductors to shield the superconducting material and to provide mechanical strength and strain relief [30]. It is beneficial to know the mechanical properties of Ag on a small scale including the power law that governs the size effect. Combined intrinsic and extrinsic size effects have been observed in single-crystal and polycrystalline Ag micropillars made through an embossing technique [13]. This study found that smaller samples with nearly single-crystalline structure— though containing dislocations—exhibit large strain bursts while larger polycrystalline samples show a less-serrated stress-strain response similar to the bulk material response. Strain bursts are caused by an avalanche of dislocations spontaneously nucleating and exiting the crystal and have been shown to occur less readily in polycrystalline materials since grain boundaries hinder the dislocation avalanche propagation [31]. Another study used MD simulations to explore the size effects in multi-faceted, Wulff-shaped nanoparticles made from different materials, including single-crystal silver for which a size effect of $\sigma_y = 26.83 d^{-0.481}$ is reported [25].

*2.2 Synthesis of silver micro- and nano-cubes*

We synthesized Ag micro/nano-cubes using a bottom-up seed-growth process [32,33] that results in a large quantity of monodispersed single-crystals (Fig.1). The 100 nm cubes are initially synthesized with near-perfect cubic geometry [32], having slightly rounded or truncated vertices (Fig.1c). These nanocubes can be grown into larger cubes by performing additional synthesis steps [33]. The first seed growth step produces nanocubes with ~400-500 nm sides, and an additional seed growth step produces microcubes (1000-2000 nm, Fig.1a). The microcubes have much sharper edges and corners with radius of curvature to edge length ratios ($r_e$) of 0.023±0.003, compared to 0.138±0.012 for the ~100 nm sized nanocubes (Fig.1(b-c)).

STEM analyses characterizes the as-synthesized cube's internal structure. Selected area diffraction (SAD) reveals a perfect fcc single-crystal structure (Fig.1d). The high-resolution STEM images show a pristine fcc structure, free of any dislocations—as in the representative image shown in Fig.1e. The primary advantages of this seed-growth process are that the synthesized Ag-cubes are dislocation-free and are not substrate bound, in contrast to other top-down synthesis methods such as focused ion beam (FIB) milling [4,5,7,8,22] and direct nanoimprinting or embossing [13]. The wide range of well-controlled sample sizes this synthesis method yields is instrumental to studying the size effects of initially dislocation-free single-crystal samples.

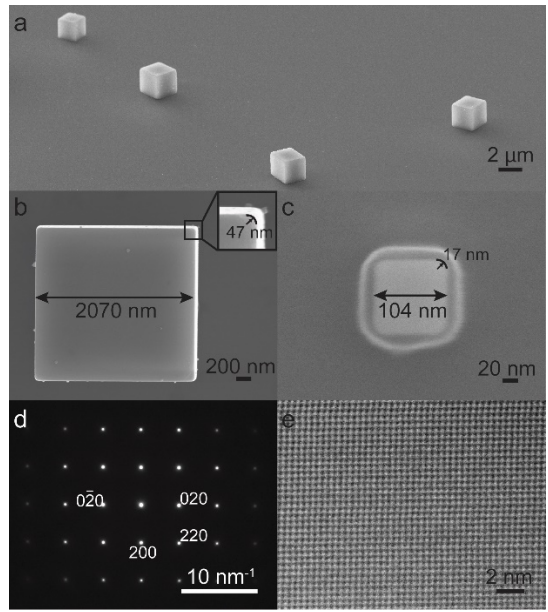

**Figure 1:** Seed-grown micro- and nano-cubes: a) scanning electron microscope (SEM) image of dispersed ~2 um cubes b) SEM image of ~2 um cube with 47 nm edge radius of curvature, c) SEM image of ~100 nm cube with 17 nm

edge radius of curvature (halo effect is from polyvinylpyrrolidone (PVP) coating), d) (001) selected area diffraction (SAD) pattern showing single-crystal fcc structure, e) HRSTEM image along [001] showing dislocation-free atomic structure.

*2.3 FIB-fabricated micropillars*

To investigate how surface ion-damage from a focused ion beam (FIB) and sample geometry may affect the deformation response, we milled nanopillars with square and circular cross sections out of our large microcubes using a Zeiss Auriga FIB (20 kV $Ga^+$ beam) and performed micro-compression tests on them. The circular pillars were milled through successively lower-current mill steps of an annular shape with a final current of 50 pA to achieve a smooth surface with minimal taper (average taper=2.6 deg). Square cross section pillars (with the new side surfaces parallel to the respective parent cube surfaces) were fabricated by milling down the edges of the cube to a desired size using a 50 pA current ion beam.

*2.4 Quasi-static compression tests*

To prepare the sample for quasi-static compression testing, the ethanol solutions containing the Ag-cubes are mixed using a vortex mixer (for cubes >500nm, ~3 min) or sonicator (for ~100nm cubes, ~10 min) to deagglomerate the cubes. We recognize that mixing the solutions could potentially lead to surface damage if the cubes impact each other, but it is necessary to ensure the cubes are well-dispersed to individually test them (further discussed later). The Ag-cubes in ethanol are then drop-casted and air-dried onto a silicon substrate. The thin (2-5 nm [33]) polyvinylpyrrolidone (PVP) coating that was used to stabilize the Ag-cubes in ethanol also acts as a buffer between the indenter and sample as well as the sample and substrate. This interface allows for more uniform loading of the sample by reducing the asperity-induced stress concentrations at the interfaces. The binding energy of the PVP to Ag could potentially have an effect on the yield strength [34], but this effect is expected to be negligible compared to the size effects studied here.

We used a Hysitron PI-85 pico-indenter equipped with a flat punch diamond tip in an SEM to perform uniaxial compression tests on the Ag-cubes with *in situ* visualization of the deformation. All quasi-static compression tests were performed in displacement-controlled mode at 0.01 s$^{-1}$ strain rate. Since the PI-85 is inherently load controlled, a PID feedback loop is necessary to use displacement-controlled mode, and the respective gains need to be tuned for our specific conditions (see SI, section 1 for more info). Properly tuned gains ensure that the indenter progresses at the prescribed displacement rate. However, if unexpected dramatic instabilities occur in the material, the actual indenter movement may deviate from the prescribed path because of inherent feedback response lag. Consequences of this will be discussed in the later sections. Prior to loading the sample, the flat punch is kept out-of-contact with the sample to maintain the pristine structure of the cube.

*2.5 Molecular statics (MS) simulations*

We use MS simulations to study the quasi-static compression of Ag nanoparticles and the effects of structural and geometric changes such as geometry, edge roundedness, vacancy concentration, and sample size. The material (Ag) was modeled with the Embedded Atom Method interatomic potential developed by Williams *et al.* [35].

Similar to the experiments, two geometries are considered: cubes and circular pillars. Multiple single-crystal Ag cubes were generated by constructing the fcc lattice with an initial lattice constant $a_0$= 0.409 nm, then allowing the lattice to relax at T = 300K. Several computational cells were simulated. The simulation cell is made up of $120^3$-$240^3$ fcc unit cells containing approximately 6 to 62 million atoms ($L_0 = 20 - 240\ a_0$). The cylindrical nanopillar was generated with radius of 32 nm and a length of around 64 nm containing 16 million atoms. Both geometries were generated with the [100]-direction coinciding with the x-direction along which the compressive load was applied.

Additionally, to investigate the effects of edge roundness, rounded edges were generated with a python script that removed atoms from the perfect cube. Geometries were generated with a range of cube lengths ($7.5 < L_0 < 60$ nm) and edge radius sizes ($0 < R < 12$ nm), to achieve edge roundness ratios ($r_e = R/L_0$) of 0-20%. A schematic cross-section of the rounded particle can be seen in Figure 2a, showing the rounded corners and labeled dimensions. The corners have a spherical surface made of discrete (111) close-pack planes of atoms as shown in Figure 2b. We also studied the effects of vacancy concentration by randomly removing individual atoms from the perfect crystal.

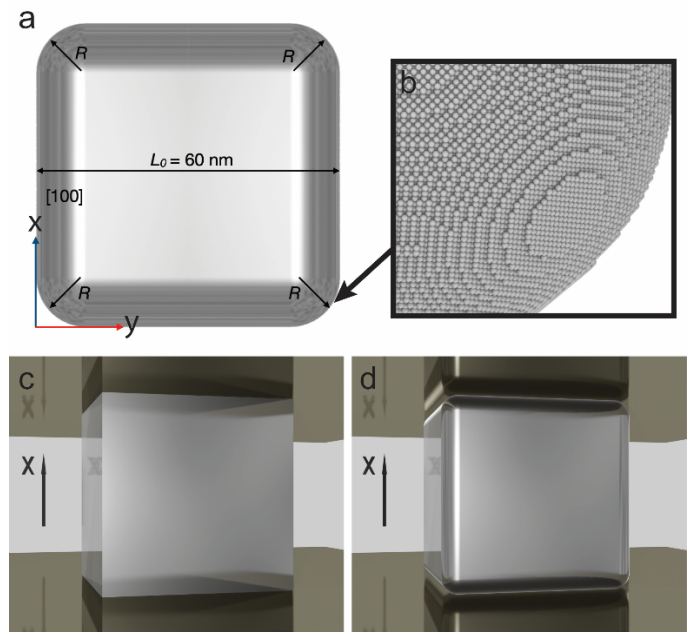

**Figure 2:** Molecular statics simulation geometry: a) <100> view of cube with rounded edges, edge radius ($R$) and side length ($L_0$) labeled, b) atomic steps forming the rounded corner, c) compression of a perfect cube with sharp edges and corners, d) compression of a cube with rounded edges and corners.

Once the geometry is generated, strain-controlled compression of the particle is performed by applying a homogeneous deformation gradient ($F$) to the atoms along x-direction. The homogeneous deformation gradient is defined as:

$$F = \begin{bmatrix} 1+\lambda & 0 & 0 \\ 0 & 1 & 0 \\ 0 & 0 & 1 \end{bmatrix}.$$

At each indentation step, the strain is increased by equal increments controlling $\lambda$. After the homogeneous deformation gradient is applied, two thin layers ($\sim 2a_0$) at the bottom and top of the particles are restricted to move in the x-direction to generate a strain-controlled compression. The positions of the atoms are optimized to minimize the potential energy of the system using the Polak-Ribière non-linear conjugate gradient (NLCG) method [36]. The tolerances were set to $10^{-10}$ eV and $10^{-10}$ eV·Å$^{-1}$ for energy and force convergence, respectively. The indentation stress was monitored using the virial stress tensor, computed using the expression

$$\sigma_{ij} = \frac{1}{2\Omega} \sum_{k=1}^{N} \sum_{\substack{l=1 \\ k \neq l}}^{N} (r_i^l - r_i^k) f_j^{kl},$$

where $\Omega$ is the volume of the particle, $r_i^l$ is the $i$-th component of the position vector of atom $l$, and $f_j^{kl}$ is the $j$-th component of the force vector applied on $k$-th atom by the $l$-th atom. $N$ is the total number of atoms in the simulation. To visualize the yield stress, we use the von Mises equivalent stress as defined in continuum mechanics. Dislocations were identified using the Dislocation Analysis (DXA) [37] implemented in open visualization tool OVITO [38].

## 3. Results

*3.1 Deformation characteristics and deformed microstructure*

Figure 3 shows an example of the quasi-static compression response of an Ag microcube (effective diameter: D=2298 nm). (See supplementary video 1 for the deformation response captured during in situ SEM compression.) The sample exhibits linear elastic behavior up to a yield strength of ~550 MPa (point 2 in Fig.3a), which is over 9x higher than the bulk yield strength of Ag ($\sigma_{bulk\ Ag}$~60MPa [30]). The jump in strain between points (2) and (4) on the stress-strain curve (Fig.3a) corresponds to a spontaneous strain-burst by almost 50% of its original height within a 20 ms interval. This instantaneous strain burst is caused by dislocation avalanches that rapidly propagate through the crystal and exit the sample surface, forming surface slip steps as seen in the SEM image (Fig.3b). Large strain bursts such as this are due to simultaneous activation of multiple slip systems. Under a perfect <100> loading condition, Schmid factor analysis predicts slip is equally likely to occur on eight different slip systems (see SI, section 2 for details). Several of these slip systems may have been activated simultaneously in this sample to accommodate the massive strain burst.

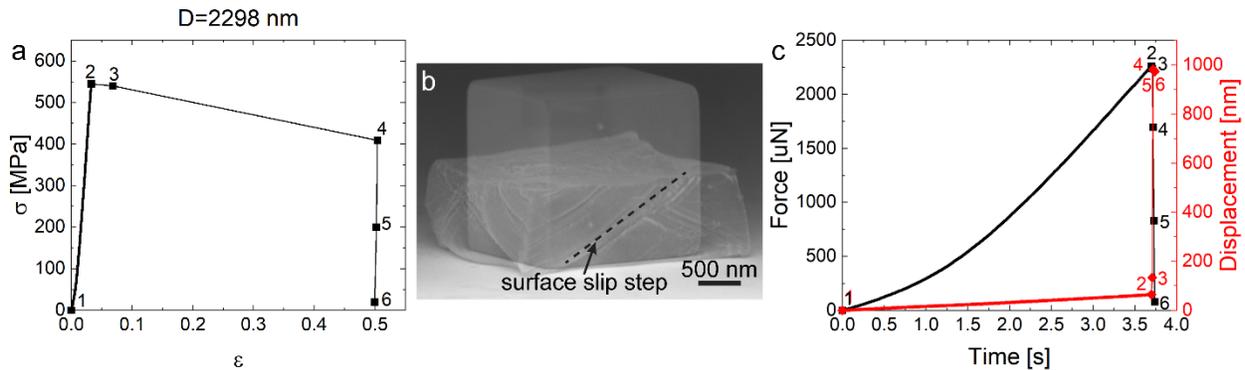

**Figure 3:** Compression response of an Ag microcube loaded along [100]: a) stress-strain response with important features labeled, b) semi-transparent SEM image of microcube before compression overlaid on SEM image taken after strain burst; notice multiple surface slip steps apparent in the deformed cube, c) force-time and displacement-time curves with important features corresponding to (a) labeled.

The strain burst can be studied further from the force-time and displacement-time plots (Fig.3c). The indenter moves at the prescribed constant displacement rate from point (1) to point (2) (corresponding to the linear elastic regime). During the strain burst, the indenter moves quickly with the sample as can be seen from the displacement-time data (points 2-4). We see from the force-time data that the force decreases rapidly during the strain burst (points 2-4) but does not immediately go to zero until point (6). This suggests that the indenter is still doing work on the cube and the entire deformation during the strain burst may not be solely due to the material instability. Ideally, the indenter would not follow the cube and would maintain the prescribed displacement rate. However, since the system is inherently load controlled, a feedback loop is required to run in pseudo-displacement-controlled mode. Consequently, during the dramatic instability of the strain burst, the indenter deviates from the prescribed displacement rate and there is a certain time lag until the feedback loop corrects the indenter position to the subsequent step to where it was before the burst.

Using TEM, we investigated the post-compression microstructure of a sample immediately after the initial large strain-burst, before further compression (Fig.4(a,c)). A thin (<100nm) <100> lamella was prepared using a focused ion beam (FEI Helios NanoLab 660 DualBeam system). The specimen surfaces were polished with a low energy $Ga^+$ beam (2 keV) to minimize FIB-induced surface damage. An FEI Titan Themis$^3$ TEM was used at an operating voltage of 300 keV to examine the deformed microstructure. A high dislocation content is visible in the bright-field TEM images (Fig.4(d,e)). The dark contrast bands intersecting in the sample's interior (Fig.4e) are avalanches of dislocations that moved through the crystal on the {111} planes. The SAD obtained across the cross-section (representative SAD in Fig.4b) reveals that the structure remains single-crystal even after being compressed by over 20%, albeit with a high-density of dislocation loops. Remarkably, this deformed microstructure is very different from that achieved during high-velocity impact of this same type of sample which was deformed to comparable strains [39].

Instead of remaining single-crystal, impacted microcubes contain a polycrystalline microstructure resulting from dynamic recrystallization and phase transformations [39,40].

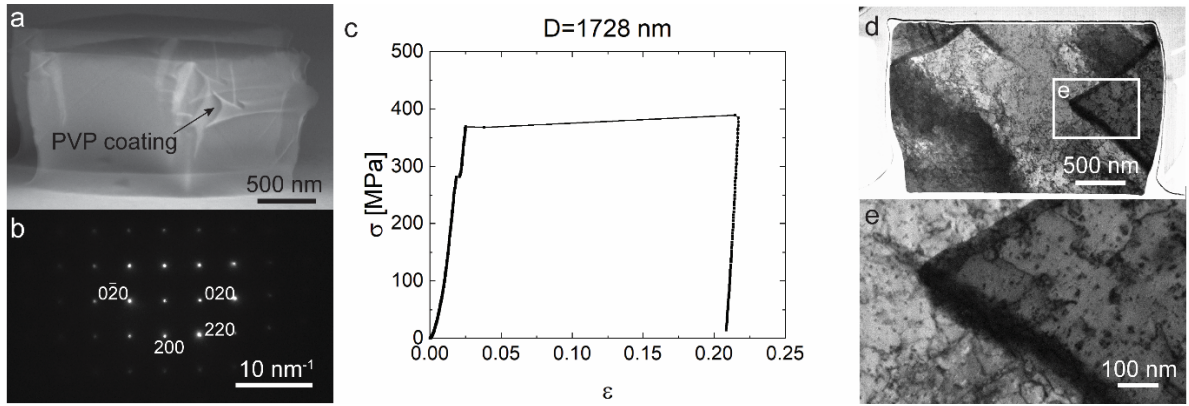

**Figure 4:** Post-compression microstructure immediately after the massive strain-burst: a) SEM image of a sample compressed until the massive strain-burst with semi-transparent image of the sample before compression overlaid, b) SAD pattern showing single-crystal structure was retained, c) deformation response of the compressed cube in (a), d) bright-field TEM image of a cross sectional lamellae (<100 nm thick, viewed along the <100> direction) of the compressed cube in (a) showing large dislocation density throughout sample, e) magnified view of the region identified in (d) showing dislocation avalanches on {111} planes.

*3.2 Stress-strain responses of cubes across sizes*

We performed quasi-static compression tests on 25 Ag-cubes ranging in size from about 100 nm to 2000 nm. The smallest ~100 nm cubes yield at phenomenal strengths, up to a quarter of the theoretical strength of silver loaded in the <100> direction (SI, section 6). The massive strain bursts were not witnessed in all samples (Fig.5b); out of the 25 cubes tested, nine did not exhibit a massive strain burst (see SI, section 3 for SEM images and stress-strain curves for all samples tested). Instead, these samples exhibit a strain hardening behavior immediately after the initial linear elastic regime. Post-compression SEM images and *in situ* SEM images reveal the samples deform through gradual slip on multiple planes in contrast to the strain-burst samples in which slips in multiple slip systems occur simultaneously.

The absence of a massive strain burst suggests conditions were not adequate to induce simultaneous slip on multiple slip systems and parallel planes. This could be due to an imperfect sample or non-uniform loading, resulting from a misaligned indenter or small asperities on the sample, indenter, or substrate surfaces. Although, we carefully ensured that the indenter was well aligned with the sample loading axis prior to loading. Some SEM images of the samples before compression show key differences in the samples which exhibited a strain burst and those that did not (Fig.5(d, e), SI section 3). For example, a pristine sample surface and sharp edges and corners are seen in the SEM image of a sample (D=2174 nm) which exhibited a strain burst (Fig.5d). In contrast, jagged edges are apparent in the SEM image of a similarly sized sample which did not exhibit a strain burst (Fig.5e). The jagged edges could have formed during synthesis or resulted from inter-particle collisions during vortex mixing. These findings suggest that surface and edge defects can significantly alter the deformation response of the cube.

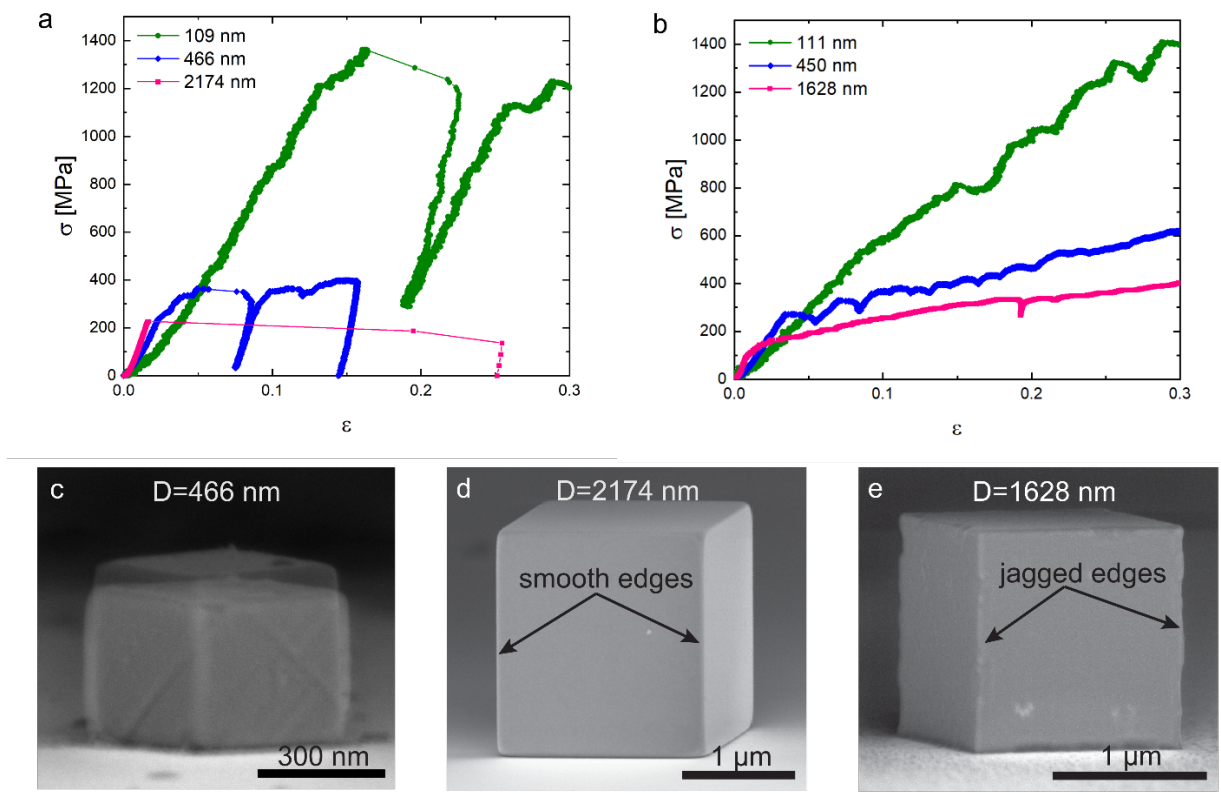

**Figure 5:** Stress-strain responses: a) stress-strain curves of three representative samples showing a strain burst, b) stress-strain curves of three representative samples which did not exhibit a large strain burst, c) semi-transparent SEM image of nanocube (D=466 nm) before compression overlaid on SEM image taken after strain burst, d) SEM image of cube which exhibited a strain burst (D=2174 nm); cube has smooth edges and surfaces, e) SEM image of cube which did not exhibit a strain burst (D=1628 nm); cube has jagged edges.

The strain-burst and no-strain-burst responses are shown in representative stress-strain curves for various sample sizes (Fig.5(a, b)). An effective diameter, representing the cube size, was calculated by equating the cross-sectional area of the cube to a circular area. A size effect on yield strength—where smaller samples are stronger—is evident in both the strain-burst and no-strain-burst sample sets. Considering the strain-burst data (Fig.5a), the smallest sample shown (D=109 nm, $\sigma_y$=1353 MPa $\approx$ 23x $\sigma_{bulk\ Ag}$) yielded at a stress ~6x that of the largest sample (D=2174 nm, $\sigma_y$=225 MPa $\approx$ 4x $\sigma_{bulk\ Ag}$)). Similarly, the smallest sample (D=111 nm, $\sigma_y$=570 MPa $\approx$ 10x $\sigma_{bulk\ Ag}$)) in the no-strain-burst set (Fig.5b) yielded at a strength over 4x that of the largest sample (D=1628 nm, $\sigma_y$=132 MPa $\approx$ 2x $\sigma_{bulk\ Ag}$)). Comparing the two sample sets reveals that samples which exhibit a strain burst yield at much higher strengths than those that do not.

Clear distinctions emerge in the responses of the different sized cubes when further analyzing the stress-strain curves for the samples that exhibited a strain burst (Fig.5a; SI, section 3.1). Notice that the stress in the smallest cubes (Fig.5a (D=109 nm); ~100 nm cubes shown in SI, section 3.1) returns to a lower yield strength after the large strain burst, while the larger sized cubes (Fig.5a (D=466nm); >400 nm cubes shown in SI, section 3.1) usually exhibit a similar or even higher yield strength upon reloading as the yield strength before the strain burst. Considering the cubes to be initially dislocation-free, one would expect an extremely high initial yield strength required to spontaneously nucleate dislocations. Because not all the nucleated dislocations exit the crystal to form surface slip steps during the strain burst (as seen in TEM, Fig.4), there is a high concentration of dislocations that could be more easily activated upon reloading. Since it is energetically easier to activate existing dislocations than nucleate new dislocations, a lower yield strength in reloading would be expected as is the case for the ~100 nm sized samples. The larger cubes do not follow this trend, exhibiting just as high of a yield strength in reloading as the initial yielding. This suggests a different dislocation structure is present, possibly containing more sessile dislocations or dislocation pile ups, making continued deformation as energetically difficult as the initial yielding. Low stacking fault materials, such as silver, more easily exhibit dislocation pile-up [41].

Another observed effect of size is the apparent differences in the strain burst magnitudes (defined as the strain difference achieved during the strain burst). The largest sample shown (D=2174 nm, Fig.5a) dramatically compresses by nearly 24% of its original height, while the smallest sample shown (D=109 nm, Fig.5a) compresses by only around 2% of its original height during the strain burst. A more in-depth discussion of this can be found in the supplementary information (SI, section 4). However, because of the aforementioned instrument limitations with being inherently load controlled, we cannot make conclusions from this observation.

*3.3 Deformation characteristics of FIB-fabricated micropillars*

To investigate the influence of potential FIB-induced defects and sample geometry, FIB-fabricated square and circular pillars were tested. Compression tests were performed on ten square micropillars and seventeen circular micropillars ranging in size from ~300 nm to 2 µm. Representative stress-strain curves are shown in Figure 6c. None of the micropillars exhibited a massive strain burst upon yield as was the case for the pristine synthesized cubes. However, there are many small strain bursts as evident in the jagged curve post-yield. All samples show a hardening behavior; however, the square pillars exhibit a higher hardening rate than the circular pillars. *In situ* SEM images taken during the compression tests show that the square pillar's faces bulge out (Fig.6(a-b)), while the circular pillar also bulges around its circumference but with its deformation localized at the top of the pillar closest to the indenter tip (Fig.6(d-e)). There could be several reasons for this observed difference: the distinct cross-sectional geometries and taper angle provide differing stress profiles, the contact conditions between the sample-indenter and sample-substrate differ under large plastic deformations, and aspect ratio or surface defects could create further asymmetries in the stress profile.

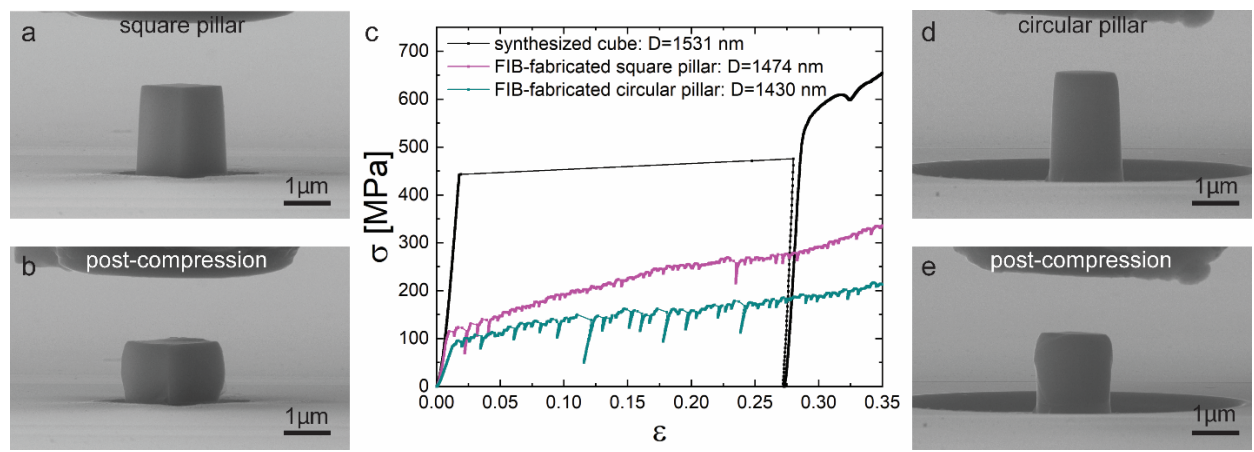

**Figure 6:** Effect of FIB-milling and geometry on deformation response: a-b) SEM images of uncompressed (a) and compressed (b) 1474 nm effective diameter square pillar with corresponding stress-strain curve in (c), c) representative stress-strain curves of compression tests on FIB-fabricated Ag micropillars with square (pink) and circular (teal) cross sections; a stress-strain curve for a synthesized cube which exhibited a strain burst is also included for comparison (black), d-e) SEM images of uncompressed (d) and compressed (e) 1430 nm diameter circular pillar with corresponding stress-strain curve in (c).

Comparing the stress-strain curves of the FIB-fabricated pillars to a similarly sized synthesized cube (Fig.6c), it is apparent that the synthesized cube yields at a drastically higher strength (~4x) than the FIB-fabricated pillars. Additionally, the FIB-fabricated pillars do not exhibit massive strain bursts like the as-synthesized cube. These differences in the stress-strain curves suggest that defects were created as a result of the $Ga^+$ ion-bombardment during FIB milling. Studies have shown that defects such as vacancies, interstitials, and dislocations can be induced as a result of the ion-bombardment during FIB milling and even when imaging with low ion doses [12,22,42]. Without the need to spontaneously nucleate dislocations, samples exhibit lower yield strength, and the dramatic strain burst does not occur.

*3.4 MS simulations: the effects of geometry, vacancy concentration, and edge roundness*

Since a size effect in initially dislocation-free samples must be dependent on the initial dislocation nucleation mechanisms, we used MS simulations to investigate what defects (other than dislocations) might contribute to differing dislocation nucleation mechanisms. We investigated three key parameters in detail: the external geometry (cylindrical vs cubic), the vacancy concentration, and the edge roundness. MS simulations of various sized perfect fcc Ag nanocubes without any of these defects do not exhibit a size effect, as one would expect with an atomically perfect fcc cubic sample. We investigated how introducing some of these defects could influence the dislocation nucleation mechanisms, and thus the size effect.

*3.4.1 Dislocation evolution in square and circular pillars*

We simulated both single-crystal cubes and cylinders to investigate the effects of geometry on dislocation nucleation and evolution. Figure 7 shows examples of dislocation evolutions in the single-crystal cube (Fig.7a) and cylinder (Fig.7b) (see supplementary videos 2-5 for the dislocation evolution in each sample; colors correspond to dislocation types as follows: blue-perfect, green-Shockley, turquoise-Frank, purple-stair rod, yellow-Hirth, red-other). The atoms that have a local structure different than fcc are shown, while atoms in an fcc structure are removed for clear visualization of dislocations and stacking faults. The first set of snapshots in Figure 7 show the initial dislocation nucleation events in the cube and cylinder. Dislocation nucleation occurs at all eight vertices of the cube, and dislocations propagate from each corner on two different slip systems (Fig.7a, step 2). In contrast, dislocations only initially nucleate on one side of the cylinder, and each of the four initial dislocations propagate on a single slip plane (Fig.7b, step 1). In both cases, as the deformation progresses the dislocations propagate on the {111} planes and intersect with each other in the center of the crystal, forming sessile dislocations (Fig.7a (step 5-7), Fig.7b (step 7)). In the cylinder, dislocations then emerge from the intersection lines and propagate out on different slip planes (Fig.7b (step 9)). The dislocation evolution continues in this fashion, while new dislocations are nucleated from the opposite end of the cylinder (Fig.7b (step 14)). The evolved dislocation networks clearly depend on the initial dislocation nucleation: a very symmetric dislocation network results from the symmetric dislocation nucleation at all eight vertices in the cube (Fig.7a (steps 44, 2)), while the asymmetric initial dislocation nucleation in the cylinder results in an asymmetric evolved dislocation network (Fig.7b (steps 1, 24)).

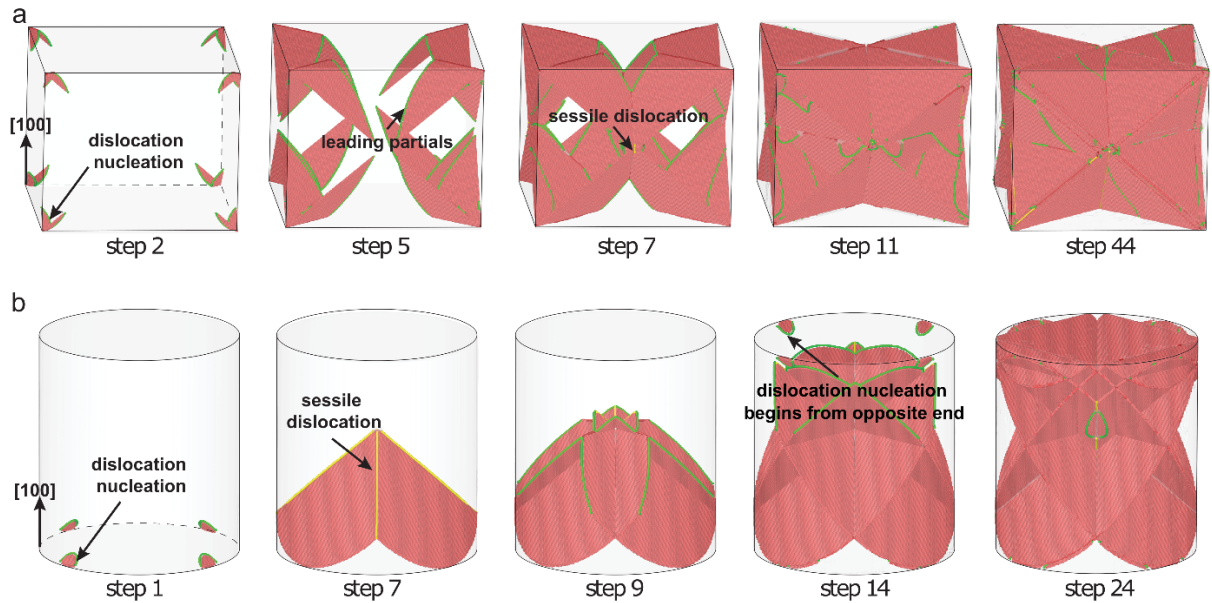

**Figure 7:** Example dislocation evolution paths in simulated compressed cube (a) and cylinder (b). [100] direction corresponds to loading direction. Atoms in fcc arrangement removed to highlight dislocations and stacking faults.

*3.4.2 Vacancy concentration and edge roundedness*

We also studied how vacancy conglomerations affect the nanocrystal plasticity, as our synthesized Ag cubes are dislocation-free, but contain point defects such as vacancies. Vacancies are intrinsic imperfections of the sample as opposed to an extrinsic shape-related imperfection that breaks the symmetry of the sample and can lead to individual strain and stress concentrations in localized regions of the sample. Vacancies were generated in the cubic sample by randomly selecting atoms and removing them from the sample. Thus, while some of these sites can lie on the surface of the sample, most of them are in the bulk of the sample. We selected samples with 0.5 and 1% vacancy concentrations. These values are well above the equilibrium vacancy concentration at room temperature ($\sim 10^{-20}$, SI section 7). Previous studies have shown that the formation energy near the surface of these samples can be reduced considerably from the bulk [43,44], and that the $Ga^+$ bombardment can generate numerous vacancies [12].

Figure 8(a-b) shows the stress-strain curves for a sample of length $L_0 = 60$ nm with different levels of vacancy concentration and edge roundness. Focusing on Figure 8a, we see the response when vacancies are added but the cube edges are kept sharp. The vacancy concentration has a drastic effect on the stress evolution in comparison with the pristine sample (p=0.0%). When there is a non-zero vacancy concentration in the sample, the peak stress reached is about ~2.25 GPa, a reduction of about ~0.5 GPa with respect to the pristine sample. When the sample has rounded edges of size $r_e = 5\%$ (Fig.8b), the vacancy concentration effect remains unchanged, as an increase in vacancy concentration from 0 to 0.5% results in a ~0.25 GPa reduction in peak stress. However, rounding the edges has a greater effect on the samples without vacancies than samples with a 0.5% vacancy concentration, as the peak stress is reduced by ~0.5 GPa when rounding the sample edges without vacancies, but remains approximately the same in the sample with 0.5% vacancy concentration. This finding suggests that the intrinsic imperfection of vacancy concentration (or vacancy conglomerations) may have more control over nanocrystal plasticity than the extrinsic imperfections in the crystal structure caused by rounded cube edges. In agreement with the previous cases, the flow stress remained ~1.75 GPa, even in the case of vacancy concentration and rounded edges combined.

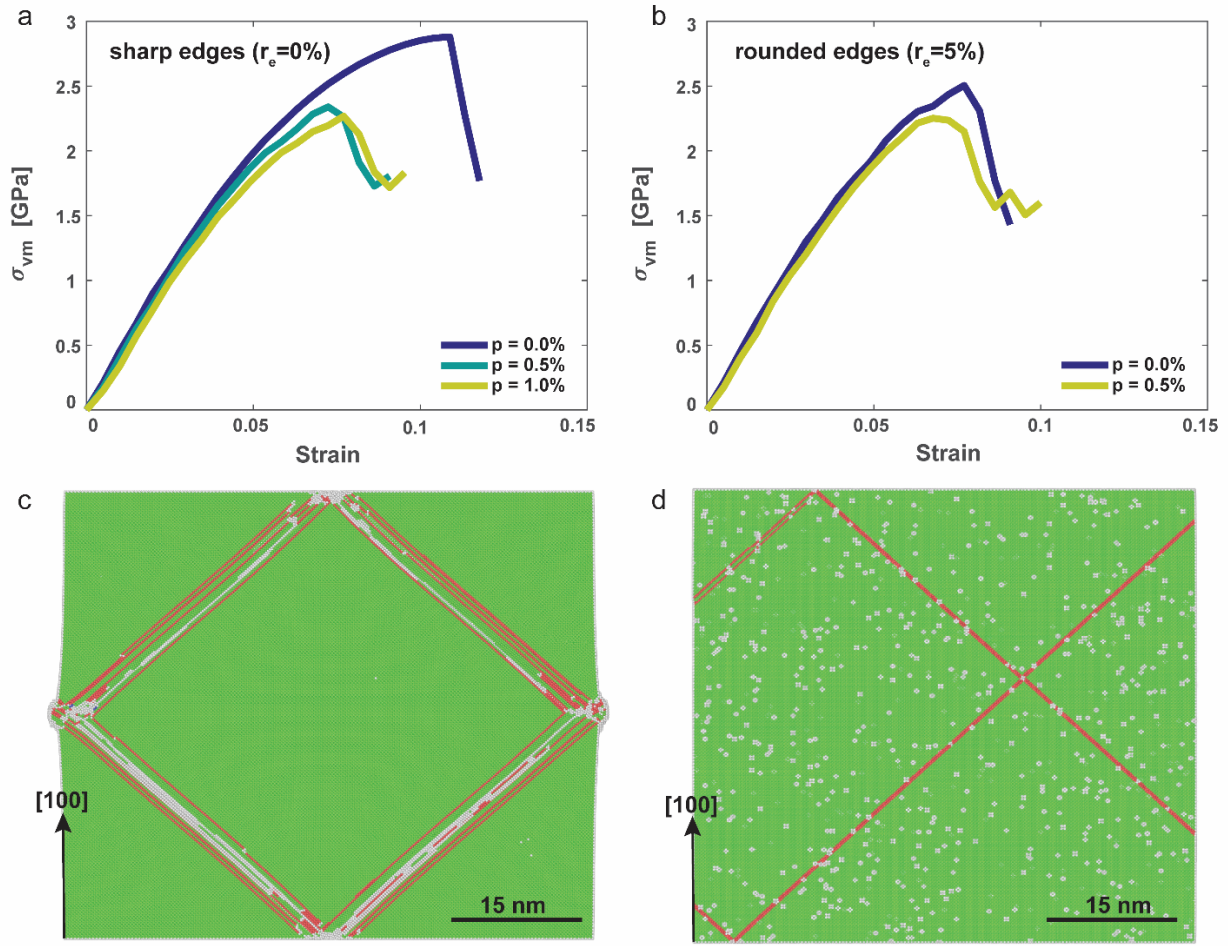

**Figure 8:** Effects of vacancy concentration: a-b) stress-strain responses for a particle with size $L_0 = 60$ nm with different vacancy concentrations and edge roundness, a) no edge roundness, and b) edge roundness ratio $r_e = 5\%$, c-d) (010) section cuts showing dislocation emission for a cube of size $L_0 = 60$ nm with sharp edges and vacancy concentrations of p=0% (c) and p=0.5% (d); atoms color coded as green: fcc, red: stacking fault, white: vacancy or atoms with no structure.

The effect of vacancy concentration on dislocation nucleation can be observed in Figure 8(c-d), where the dislocation emission in samples with no vacancies (Fig.8c) and a vacancy concentration of $p = 0.5\%$ (Fig.8d) are shown. In the case of the pristine sample shown in Figure 8c, the dislocations are initially emitted from the corners of the sample (mid-cross section shows evolved dislocation structure like in Figure 7a, step 44). On the other hand, when vacancies were placed in the sample (Figure 8d), the dislocations are now nucleated randomly in the cube. This emission is controlled by the distribution of the defect network, which breaks the symmetry observed for the pristine case. Similar behavior was observed for other vacancy concentrations.

*3.4.3 Edge roundness and sample size*

To investigate the effect of edge roundness on strength, we simulated the compression of nanocubes with a side length of $L_0 = 60$ nm and different edge roundness ratios of $r_e = R/L_0 = 0 - 20\%$ (see geometry in Figures 2a, 9a). The evolution of von Mises stress with respect to strain is shown in Figure 9b. As is evident in the figure, edge roundness has a detrimental effect on the peak strength of the particles. Note that for a small edge roundness ratio ($r_e = 1\%$), the stress-strain curve does not deviate much from that of the perfect cube ($r_e = 0\%$). However, when the edge roundness is increased further to $r_e = 5 - 20\%$, we observe a considerable drop in the peak stress from ~3 GPa to values between 2.5-1.7 GPa. However, the flow stress remains approximately constant at around ~1.75 GPa, as can be seen for the case of $r_e = 20\%$. Interestingly, the critical strain at which the peak stress is reached also decreases with edge roundness.

We observe that the decrease in peak stress and strain is related to early nucleation events seen in the rounded particles (Fig.9(a,c)). This is also evident in the stress-strain plot (Fig.9b), where the stress evolution has some fluctuations at early strains for large edge roundness ratios. In order to have a clear understanding of the mechanisms, we investigated the first dislocation nucleation event and its evolution. Figure 9a shows a (110) section through the middle of a $L_0 = 60$ nm particle, with $r_e = 20\%$ at ~3% strain. At this level of strain, the sample has not yet emitted any dislocations; however, there are elastic strain localizations near the intersections between the rounded edges and flat surfaces of the particles. These strain localizations coincide with the locations of the first dislocation nucleation events. This can be seen in Figure 9c where four planes of atoms in stacking fault structure are seen in red. The stacking fault planes indicate that several Shockley partial dislocations have been emitted from these points.

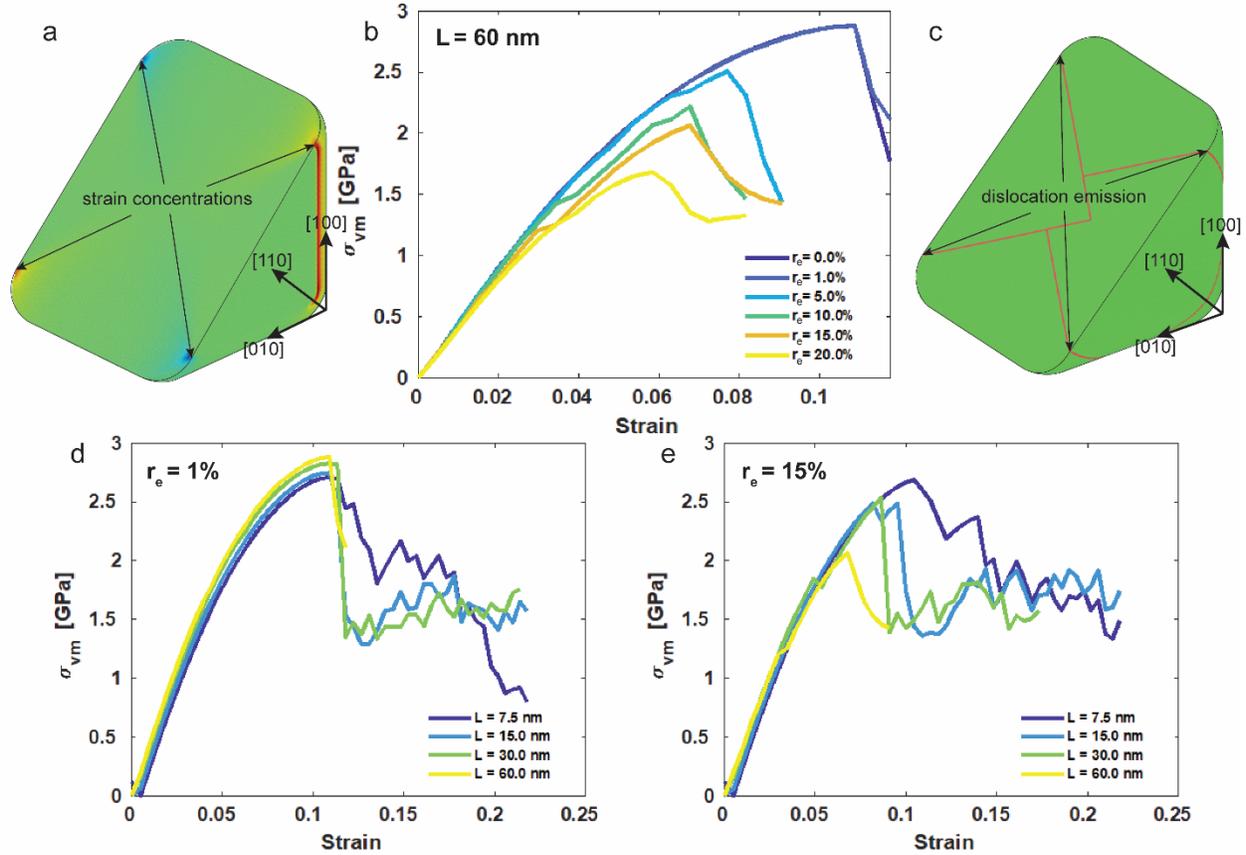

**Figure 9:** Effects of roundness: a) (110) section cut of cube with rounded edges showing elastic strain localization before first dislocation nucleation event, b) stress-strain responses for a particle size of $L_0 = 60$ nm with edge roundness ratios ranging from $r_e = 0\% - 20\%$, c) stacking faults in red, showing path of partial dislocations nucleated from corners near rounded edges, d-e) stress-strain responses of different sized particles with edge roundness ratios of 1% (d) and 15% (e).

This dislocation nucleation mechanism explains the observed trend of reduction in peak stress with increased edge roundness. The rounded edges cause local atomic defects in the perfect cubic geometry comprised of an fcc atomic structure, as can be seen from the atomic-layer surface steps which constitute the rounded geometry (Fig.2b). These surface defects are locations of strain and stress localizations, which are favorable sites for dislocation nucleation. As more rounded edges are simply longer surface defects, it will make dislocation nucleation easier, initiating plasticity at lower stresses and strains. It is important to note that this observation contradicts some other recent studies on the effects of edge roundedness on nanocrystal plasticity [23,24,27]. Amodeo and Lizoul performed MD simulations on $Ni_3Al$ nanoparticles ranging from a perfect cube to a perfect sphere by progressively rounding the cube edges [27]. Sharma *et al.* performed MD simulations on Wulff shaped Ni nanoparticles with varying degrees of edge roundness [23], and experiments and MD simulations on rounded and faceted Mo nanoparticles [24]. Each found the opposite trend—the strength increases with increased roundedness. They take a continuum perspective, arguing that the rounded edges and corners would reduce stress concentrations present at sharp corners. A few key differences in the simulation methods are noteworthy to understand the differences in results. First, these studies [23,24,27] use molecular dynamics (MD) simulations with high strain rates on the order of $10^8$ $s^{-1}$, while our molecular statics (MS) approach provides a quasi-static simulation, comparable to experiments. Furthermore, the methods used to calculate the stress differ; these three studies define an engineering stress as the force divided by the initial top surface area of the particle in contact with the indenter. By using this approach, the stress could artificially be inflated for rounded particles since the contact area is reduced as the particle roundness increases. Amodeo and Lizoul [27] recognize this and admit that use of the mid-height area instead would reverse the observed trend—where the strength would reduce with increasing roundedness. To eliminate this confusion in literature and to obtain an accurate result, we use the von Mises stress computed from the virial stress tensor.

Next, we investigate the effects of edge roundness for different sample sizes. In order to do so, we varied the sample size between $7.5 < L_0 < 60$ nm, and kept the edge radius to a constant value of $R = 0.01L_0$ and $R = 0.15L_0$. For $R = 0.01L_0$ (Fig.9d), we see a slight decrease in peak stress as the sample size is reduced—although all values are closer to ~2.75 GPa with minimal deviation. This suggests that when the edge radius is small in comparison to the size of the sample, there is not much effect. However, when the edge roundness ratio is increased to 15%, a clear size effect emerges where the peak stress decreases with increasing sample size (Fig.9e). The dislocation nucleation mechanisms described previously explain this behavior: strain localizations emerge at the intersections of the rounded edge and the flat faces of the sample which become preferential sites for dislocation nucleation. Increasing the sample size while maintaining the same edge roundness ratio results in larger surface defects, making dislocation nucleation easier. Remarkably, the flow stress also remains ~1.75 GPa for all samples, in good agreement with other cases.

## 4. Discussion

### 4.1 Size effect on yield strength

To quantitatively study the size effect on yield strength, we plot the normalized yield shear strengths versus the normalized sample sizes and describe the size effect for each sample set by a power law relation (Fig.10). The normalized data allows for comparison to literature with different loading geometries and materials. The yield strength for data without an obvious yield point is obtained by an offset strain of 0.005. The peak stress is used for the data from the MS simulations in Fig.9e. The measured yield strengths are resolved onto the {111}⟨110⟩ slip systems of the fcc lattice and then normalized by the shear modulus of Ag on {111} planes (see SI, section 5 for further details). The sample sizes are normalized by the magnitude of the Burger's vector. In accordance with common representation of size effect, a power-law relation is fit to each data set: $\sigma_y = A(D/b)^{-m}$. In addition, the universal power law relation for fcc metal micropillars [8] and the power law for Ag nanoparticles obtained by MD simulations [25] are included.

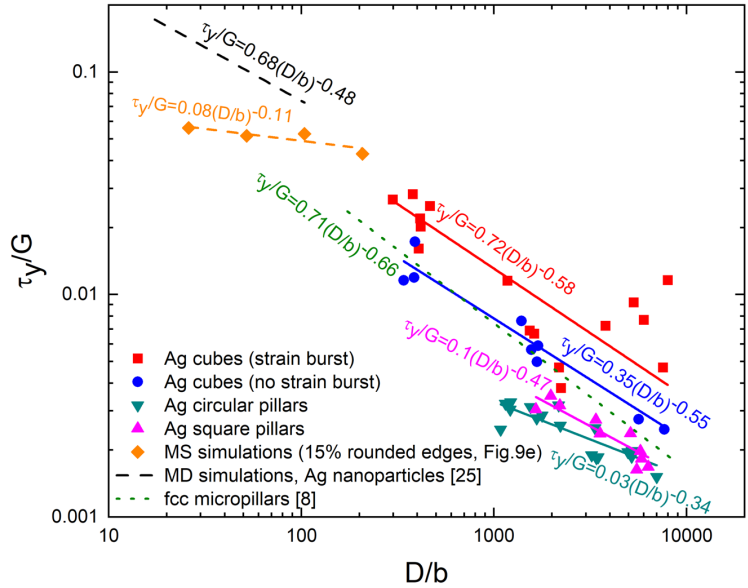

**Figure 10:** Size effect on yield strength: log-log plot of normalized shear yield strengths versus normalized sample sizes; power law fits of our experimental data: Ag strain-burst (red), no-strain-burst (blue), FIB-fabricated circular pillar (teal), and FIB-fabricated square pillar (pink) sample sets, and our MS simulations of samples with 15% edge roundness from Fig.9e (orange) compared to data from literature (modeled Ag nanoparticles [25] (black), compilation of fcc micropillars [8] (green)).

All sample sets exhibit a size effect on strength. Overall, all of our experimental data sets have a similar degree of size effect (slope: exponent of (D/b)) and are similar to values found in literature. The as-synthesized cubes (red and blue) have a slightly stronger size effect than the FIB-fabricated pillars (teal and pink). The stochastic nature of the "strain-burst" samples' deformation (red) compared to other samples' response is also evident in Fig.10. Our FIB-fabricated square pillars (pink) have a slightly larger degree of size effect (slope =-0.47) than the circular pillars (teal, slope=-0.34). The slope for the square pillars is more similar to the Ag microcubes (slope: -0.58 and -0.55 for strain-burst (red) and no-strain-burst (blue) samples, respectively) and the MD simulated Ag nanoparticles (black dashed line, slope=-0.48) [25], which might suggest the faceted geometry could play a role in the size effect. We also see a less pronounced, but clear size effect in the MS data from the cubes with 15% rounded edges (orange). This size effect originates from the defects caused by removing the edge and corner atoms. These MS simulations, in addition to the comparison of the cube and pillar data from experiments, confirms that the external geometry influences the size effect.

*4.2 Role of initial defects*

While the size effects (slopes) of the experimental data sets are similar (Fig.10), the intercept of the fitted curves vary along the y-axis—demonstrating that each sample set achieves different strengths across sample sizes. For example, the FIB-fabricated pillars (teal and pink) yield at strengths consistently lower than those of the as-synthesized cubic samples without FIB exposure (both those that exhibited strain bursts and those that did not). We attribute the lower strength to a higher initial defect content, whether that be FIB-induced dislocations or $Ga^+$ implantation. This is consistent with the trend that the higher the initial defects, the lower the yield strength is: the MD simulations of Ag nanoparticles [25] and our MS simulations of Ag nanocubes show the highest yield strengths since they are free of any internal defects. Feruz and Mordehai explain that the size effect seen in their MD simulations is due to the faceted particle shape, which elicits differing stress gradients depending on the particle size [25]. We were also able to produce a size effect in our MS simulations by rounding the cube edges, which creates atomic-level surface defects of lengths proportional to the sample size. Our Ag microcubes which exhibited a large strain burst yield at slightly lower strength, followed by the cubes which did not exhibit a strain burst, which had more initial surface defects. Lastly, the Ag micropillars yield at even lower strengths since the FIB fabrication process introduces more defects from $Ga^+$ ion bombardment. These results show that slight differences in the initial defect structure have an immense influence on the mechanical performance of the material.

We also considered how slight differences in the initial defect structure could influence the size effect. Particularly, because our as-synthesized cubes are initially dislocation-free but still exhibit a size effect, we investigated what defects (other than dislocations) could be contributing to the size effect. The samples which exhibited a strain-burst are initially dislocation free as shown in the HRTEM image in Fig.1e. Without any preexisting dislocations, the observed size effect on yield strength must be due to differing dislocation nucleation mechanisms. Dislocations can be nucleated homogeneously or heterogeneously [45]. Heterogeneous nucleation is favored as it requires much lower levels of stress than homogeneous nucleation [23]. Dislocations are nucleated heterogeneously at areas of high stress concentration [20,23,25,26,45,46]. If the stress state is perfectly uniform—free of any stress concentrations—homogenous nucleation at the theoretical stress may occur; however, this is generally not experimentally achievable. More commonly, heterogenous nucleation occurs, driven by factors that disrupt the perfect cubic symmetry and elicit stress or strain concentrations. We have investigated several of these potential factors in this work, including the external defects such as surface asperities and edge roundedness, and internal defects such as vacancy conglomerations.

External defects such as surface asperities or rounded edges can cause stress and strain concentrations to develop at localized regions and at the sample-indenter interface, initiating early dislocation nucleation at these locations. Larger cubes may have a greater probability of surface defects, which has been used to describe the size effect in terms of a weakest-link model through Weibull statistics [17,18,22]. We have also investigated the effects of external geometry and edge roundness on the size effect, both through MS simulations and experiments. In experiments, the larger cubes have sharper edges and corners than the smallest 100 nm cubes ($r_e = 2.3\%$ for microcubes, $r_e = 13.8\%$ for 100nm cubes). Since the smallest cubes yielded at higher strengths than the larger cubes (defining the size effect), this may suggest that rounder edges correlate with dislocation nucleation at higher stresses. However, MS simulations reveal that rounding the cube edges and corners will actually have a detrimental effect on the strength, since rounded edges break the perfect cubic symmetry at the atomic scale and create strain localizations where the rounded edge meet the cube faces—prime locations for dislocation nucleation. With this understanding, we can conclude that the edge roundness ratio ($r_e = R/L_0$) may not be as important as the actual edge radius size ($R$), as the disruption of the perfect cubic atomic structure is what creates the local strain concentrations, initiating dislocation nucleation at the rounded edges.

We also investigated the effect of cross-sectional geometry by testing square and circular pillars made via FIB-milling and performing MS simulations on cubes and circular pillars. While the experimental pillars could have FIB-induced dislocations, there were distinct differences seen in the deformation responses; for example, the circular pillars have more localized deformation near the flat punch, while the square pillars exhibit symmetric bulging along the height. Our MS simulations suggests that asymmetric initial dislocation nucleation events may lead to an eventual asymmetric deformation—as is the case for the cylinder—since the evolved dislocation network depends on the symmetry of the initial dislocation nucleation. These findings reveal that the external sample geometry has considerable effect on nanocrystal plasticity.

In addition to external defects, internal defects can also break the perfect cubic symmetry and influence dislocation nucleation mechanisms. The MS simulations of cubes with varying vacancy concentrations reveal that vacancy conglomerations have a strong detrimental effect on the strength since dislocations are preferentially nucleated at the stress concentrations near the voids. Combining the effects of vacancy concentration and edge roundness, the vacancy concentration has greater control over nanocrystal plasticity than edge roundness—illuminating the coupled effects of external and internal atomic-level symmetry breakers. Additional studies into the vacancy concentrations or void generations resulting from a seed-mediated growth synthesis method could provide further validation.

## 5. Conclusions

A size effect is evident in our single-crystal Ag micro- and nano-cubes that is dependent on the initial defect-structure of the sample which elicits different deformation mechanisms. The initially dislocation-free Ag nanocubes exhibit massive strain bursts and yield at strengths up to one-fourth of the theoretical strength of silver. The size effect seen in the initially dislocation-free Ag cubes must be due to differing dislocation nucleation mechanisms. We determined that surface defects and vacancy conglomerations can break the perfect cubic symmetry and initiate heterogeneous dislocation nucleation at lower stresses. The surface defects—asperities on the sample, indenter, or substrate surfaces, or FIB-induced surface amorphization and point defects—break the cubic symmetry, resulting in stress concentrations which promote early heterogeneous nucleation of dislocations. Rounding the cube edges—although does not break the cubic symmetry in sample scale—breaks the intrinsic crystal symmetry locally and results in early dislocation nucleation from the intersection points between the rounded edge and flat cube face. Experimental and simulation evidence suggests that the actual edge radius rather than the roundness ratio ($r_e =$ $R/L_0$) matters, since the rounded edges are comprised of atomic absences in the perfect cubic lattice. Our findings provide insights into what causes the strength lower than the theoretical strength in initially dislocation-free samples and an apparent size effect on their yield strength.

**Acknowledgements**

We acknowledge the support for this research provided by the University of Wisconsin-Madison, Office of the Vice Chancellor for Research and Graduate Education with funding from the Wisconsin Alumni Research Foundation. We also acknowledge partial support from Rice University and from the National Research Foundation of Korea (NRF) grant funded by the Ministry of Education (NRF-2018R1D1A1B07044075). Support for electron microscopy at the Wisconsin Centers for Nanoscale Technology of the University of Wisconsin-Madison and the Center for Electron Microscopy of Rice University are also acknowledged. This work was performed, in part, at the Center for Integrated Nanotechnologies, an Office of Science User Facility operated for the U.S. Department of Energy (DOE) Office of Science. Los Alamos National Laboratory, an affirmative action equal opportunity employer, is managed by Triad National Security, LLC for the U.S. Department of Energy's NNSA, under contract 89233218CNA000001.